
%
%
\hsize=6.5truein
\hoffset=0truein
\vsize=8.9truein
\voffset=0truein
\font\twelverm=cmr10 scaled 1200    \font\twelvei=cmmi10 scaled 1200
\font\twelvesy=cmsy10 scaled 1200   \font\twelveex=cmex10 scaled 1200
\font\twelvebf=cmbx10 scaled 1200   \font\twelvesl=cmsl10 scaled 1200
\font\twelvett=cmtt10 scaled 1200   \font\twelveit=cmti10 scaled 1200
\skewchar\twelvei='177   \skewchar\twelvesy='60
\def\twelvepoint{\normalbaselineskip=12.4pt
  \abovedisplayskip 12.4pt plus 3pt minus 9pt
  \belowdisplayskip 12.4pt plus 3pt minus 9pt
  \abovedisplayshortskip 0pt plus 3pt
  \belowdisplayshortskip 7.2pt plus 3pt minus 4pt
  \smallskipamount=3.6pt plus1.2pt minus1.2pt
  \medskipamount=7.2pt plus2.4pt minus2.4pt
  \bigskipamount=14.4pt plus4.8pt minus4.8pt
  \def\rm{\fam0\twelverm}          \def\it{\fam\itfam\twelveit}%
  \def\sl{\fam\slfam\twelvesl}     \def\bf{\fam\bffam\twelvebf}%
  \def\mit{\fam 1}                 \def\cal{\fam 2}%
  \def\tt{\twelvett}
  \textfont0=\twelverm   \scriptfont0=\tenrm   \scriptscriptfont0=\sevenrm
  \textfont1=\twelvei    \scriptfont1=\teni    \scriptscriptfont1=\seveni
  \textfont2=\twelvesy   \scriptfont2=\tensy   \scriptscriptfont2=\sevensy
  \textfont3=\twelveex   \scriptfont3=\twelveex  \scriptscriptfont3=\twelveex
  \textfont\itfam=\twelveit
  \textfont\slfam=\twelvesl
  \textfont\bffam=\twelvebf \scriptfont\bffam=\tenbf
  \scriptscriptfont\bffam=\sevenbf
  \normalbaselines\rm}

\font\titlerm=cmr10 scaled\magstep3
\font\titleit=cmti10 scaled\magstep3

\def\titlefonts{\def\rm{\titlerm}
                \def\it{\titleit} \rm}
\font\twelvesc=cmcsc10 scaled 1200

\def\beginlinemode{\endmode
  \begingroup\parskip=0pt \obeylines\def\\{\par}\def\endmode{\par\endgroup}}
\def\beginparmode{\endmode
  \begingroup \def\endmode{\par\endgroup}}
\let\endmode=\par
{\obeylines\gdef\
{}}
\def\singlespace{\baselineskip=\normalbaselineskip}
\def\oneandahalfspace{\baselineskip=\normalbaselineskip
  \multiply\baselineskip by 3 \divide\baselineskip by 2}
\def\doublespace{\baselineskip=\normalbaselineskip \multiply\baselineskip by 2}

\newcount\firstpageno
\firstpageno=2
\footline={\ifnum\pageno<\firstpageno{\hfil}\else{\hfil\twelverm\folio\hfil}\fi}
\let\rawfootnote=\footnote		
\def\footnote#1#2{{\rm\singlespace\parindent=0pt\rawfootnote{#1}{#2}}}
\def\raggedcenter{\leftskip=4em plus 12em \rightskip=\leftskip
  \parindent=0pt \parfillskip=0pt \spaceskip=.3333em \xspaceskip=.5em
  \pretolerance=9999 \tolerance=9999
  \hyphenpenalty=9999 \exhyphenpenalty=9999 }
\parskip=\medskipamount
\twelvepoint		
\doublespace		
\overfullrule=0pt	

\def\preprintno#1{
 \rightline{\rm #1}}	

\def\title                      
  {\null\vskip 3pt plus 0.2fill
   \beginlinemode \doublespace \raggedcenter \titlefonts}

\def\author                     
  {\vskip 3pt plus 0.2fill \beginlinemode
   \singlespace \raggedcenter\twelvesc}

\def\affil			
  {\vskip 3pt plus 0.03fill \beginlinemode
   \oneandahalfspace \it \centerline}

\def\abstract			
  {\vskip 3pt plus 0.3fill \beginparmode
   \doublespace \narrower ABSTRACT: }

\def\endtitlepage		
  {\endpage			
   \body}

\def\body			
  {\beginparmode}		

\def\head#1{			
  \filbreak\vskip 0.5truein	
  {\immediate\write16{#1}
   \raggedcenter \bf\uppercase{#1}\par}
   \nobreak\vskip 0.25truein\nobreak}

\def\references
  {\head{REFERENCES}
   \frenchspacing \parindent=0pt \leftskip=0.8truecm \rightskip=0truecm
   \parskip=4pt plus 2pt \everypar{\hangindent=\parindent}}

\def\pr{\journal Phys. Rev.}

\def\zphys{\journal Z. Phys.}

\def\ptp{\journal Prog. Theor. Phys.}

\def\endreferences{\body}

\def\figurecaptions		
  {\endpage
   \beginparmode
   \head{Figure Captions}
}

\def\endpage			
  {\vfill\eject}

\def\endpaper			
  {\endmode\vfill\supereject}
\def\endit
  {\endpaper\end}

\def\frac#1#2{{\textstyle{#1 \over #2}}}
\def\ref#1{ref. #1}
\def\Ref#1{Ref. #1}


\def\sla{\raise.15ex\hbox{$/$}\kern-.57em}
\def\leaderfill{\leaders\hbox to 1em{\hss.\hss}\hfill}
\def\twiddle{\lower.9ex\rlap{$\kern-.1em\scriptstyle\sim$}}
\def\bigtwiddle{\lower1.ex\rlap{$\sim$}}
\def\gtwid{\mathrel{\raise.3ex\hbox{$>$\kern-.75em\lower1ex\hbox{$\sim$}}}}
\def\ltwid{\mathrel{\raise.3ex\hbox{$<$\kern-.75em\lower1ex\hbox{$\sim$}}}}
\def\square{\kern1pt\vbox{\hrule height 1.2pt\hbox{\vrule width 1.2pt\hskip 3pt
   \vbox{\vskip 6pt}\hskip 3pt\vrule width 0.6pt}\hrule height 0.6pt}\kern1pt}

%
%
%
\def\refstylenp{		
  \gdef\refto##1{~[##1]}				
  \gdef\r##1{~[##1]}	         			
  \gdef\refis##1{\indent\hbox to 0pt{\hss[##1]~}}     	
  \gdef\citerange##1##2##3{\cite{##1}--\setbox0=\hbox{\cite{##2}}\cite{##3}}
  \gdef\rrange##1##2##3{~[\cite{##1}--\setbox0=\hbox{\cite{##2}}\cite{##3}]}
  \gdef\journal##1, ##2, ##3,                           
    ##4,{{\sl##1} {\bf ##2} (##3) ##4}}
\def\refstylezphys{		
  \gdef\refto##1{~[##1]}				
  \gdef\r##1{~[##1]}	         			
  \gdef\refis##1{\indent\hbox to 0pt{\hss[##1]~}}     	
  \gdef\citerange##1##2##3{\cite{##1}--\setbox0=\hbox{\cite{##2}}\cite{##3}}
  \gdef\rrange##1##2##3{~[\cite{##1}--\setbox0=\hbox{\cite{##2}}\cite{##3}]}
  \gdef\journal##1, ##2, ##3,                           
    ##4,{{\sl##1} {\bf ##2}, ##4 (##3)}}
\def\refstylepr{		
  \gdef\refto##1{~[##1]}		
  \gdef\r##1{~[##1]}		        
  \gdef\refis##1{\indent\hbox to 0pt{\hss[##1]~}}	
  \gdef\citerange##1##2##3{\cite{##1}--\setbox0=\hbox{\cite{##2}}\cite{##3}}
  \gdef\rrange##1##2##3{~[\cite{##1}--\setbox0=\hbox{\cite{##2}}\cite{##3}]}
  \gdef\journal##1, ##2, ##3,                           
    ##4,{{\sl##1} {\bf ##2}, ##4 (##3)}}
\def\refstyleijmp{		
  \gdef\refto##1{$^{##1}$}				
  \gdef\r##1{$^{##1}$}	         			
  \gdef\refis##1{\indent\hbox to 0pt{\hss##1.~}}     	
  \gdef\citerange##1##2##3{\cite{##1}--\setbox0=\hbox{\cite{##2}}\cite{##3}}
  \gdef\rrange##1##2##3{$^{\cite{##1}-\setbox0=\hbox{\cite{##2}}\cite{##3}}$}
  \gdef\journal##1, ##2, ##3,                           
    ##4,{{\sl##1} {\bf ##2} (##3) ##4}}
\def\(#1){(\call{#1})}
\def\call#1{{#1}}
\def\smu
{Department of Physics, Southern Methodist University, Dallas, TX 75275}

\def\e{\epsilon}

\def\del{\partial}
\def\ha{{1\over 2}}

\def\bar#1{\overline{ #1 }}

\def\ran{\rangle}

\def\ket#1{|#1\ran}

\def\normord#1{\mathopen{\hbox{\bf:}}#1\mathclose{\hbox{\bf:}}}

\def\kd3{\delta^{(3)}}
\def\sym#1,#2{\Biggl\{ {#1}{1\over i\del^+} {#2} \Biggr\}_{\rm sym} }
\def\ssym#1,#2{\Biggl\{ {#1}{1\over (i\del^+)^2} {#2} \Biggr\}_{\rm sym} }
\refstylenp
\catcode`@=11
\newcount\tagnumber\tagnumber=0
\immediate\newwrite\eqnfile
\newif\if@qnfile\@qnfilefalse
\def\write@qn#1{}
\def\writenew@qn#1{}
\def\w@rnwrite#1{\write@qn{#1}\message{#1}}
\def\@rrwrite#1{\write@qn{#1}\errmessage{#1}}
\def\taghead#1{\gdef\t@ghead{#1}\global\tagnumber=0}
\def\t@ghead{}
\expandafter\def\csname @qnnum-3\endcsname
  {{\t@ghead\advance\tagnumber by -3\relax\number\tagnumber}}
\expandafter\def\csname @qnnum-2\endcsname
  {{\t@ghead\advance\tagnumber by -2\relax\number\tagnumber}}
\expandafter\def\csname @qnnum-1\endcsname
  {{\t@ghead\advance\tagnumber by -1\relax\number\tagnumber}}
\expandafter\def\csname @qnnum0\endcsname
  {\t@ghead\number\tagnumber}
\expandafter\def\csname @qnnum+1\endcsname
  {{\t@ghead\advance\tagnumber by 1\relax\number\tagnumber}}
\expandafter\def\csname @qnnum+2\endcsname
  {{\t@ghead\advance\tagnumber by 2\relax\number\tagnumber}}
\expandafter\def\csname @qnnum+3\endcsname
  {{\t@ghead\advance\tagnumber by 3\relax\number\tagnumber}}
\def\equationfile{%
  \@qnfiletrue\immediate\openout\eqnfile=\jobname.eqn%
  \def\write@qn##1{\if@qnfile\immediate\write\eqnfile{##1}\fi}
  \def\writenew@qn##1{\if@qnfile\immediate\write\eqnfile
    {\noexpand\tag{##1} = (\t@ghead\number\tagnumber)}\fi}
}
\def\callall#1{\xdef#1##1{#1{\noexpand\call{##1}}}}
\def\call#1{\each@rg\callr@nge{#1}}
\def\each@rg#1#2{{\let\thecsname=#1\expandafter\first@rg#2,\end,}}
\def\first@rg#1,{\thecsname{#1}\apply@rg}
\def\apply@rg#1,{\ifx\end#1\let\next=\relax%
\else,\thecsname{#1}\let\next=\apply@rg\fi\next}
\def\callr@nge#1{\calldor@nge#1-\end-}
\def\callr@ngeat#1\end-{#1}
\def\calldor@nge#1-#2-{\ifx\end#2\@qneatspace#1 %
  \else\calll@@p{#1}{#2}\callr@ngeat\fi}
\def\calll@@p#1#2{\ifnum#1>#2{\@rrwrite{Equation range #1-#2\space is bad.}
\errhelp{If you call a series of equations by the notation M-N, then M and
N must be integers, and N must be greater than or equal to M.}}\else%
 {\count0=#1\count1=#2\advance\count1
by1\relax\expandafter\@qncall\the\count0,%
  \loop\advance\count0 by1\relax%
    \ifnum\count0<\count1,\expandafter\@qncall\the\count0,%
  \repeat}\fi}
\def\@qneatspace#1#2 {\@qncall#1#2,}
\def\@qncall#1,{\ifunc@lled{#1}{\def\next{#1}\ifx\next\empty\else
  \w@rnwrite{Equation number \noexpand\(>>#1<<) has not been defined yet.}
  >>#1<<\fi}\else\csname @qnnum#1\endcsname\fi}
\let\eqnono=\eqno
\def\eqno(#1){\tag#1}
\def\tag#1$${\eqnono(\displayt@g#1 )$$}
\def\aligntag#1\endaligntag
  $${\gdef\tag##1\\{&(##1 )\cr}\eqalignno{#1\\}$$
  \gdef\tag##1$${\eqnono(\displayt@g##1 )$$}}

\def\eqalignno#1{\displ@y \tabskip\centering
  \halign to\displaywidth{\hfil$\displaystyle{##}$\tabskip\z@skip
    &$\displaystyle{{}##}$\hfil\tabskip\centering
    &\llap{$\displayt@gpar##$}\tabskip\z@skip\crcr
    #1\crcr}}
\def\displayt@gpar(#1){(\displayt@g#1 )}
\def\displayt@g#1 {\rm\ifunc@lled{#1}\global\advance\tagnumber by1
        {\def\next{#1}\ifx\next\empty\else\expandafter
        \xdef\csname @qnnum#1\endcsname{\t@ghead\number\tagnumber}\fi}%
  \writenew@qn{#1}\t@ghead\number\tagnumber\else
        {\edef\next{\t@ghead\number\tagnumber}%
        \expandafter\ifx\csname @qnnum#1\endcsname\next\else
        \w@rnwrite{Equation \noexpand\tag{#1} is a duplicate number.}\fi}%
  \csname @qnnum#1\endcsname\fi}
\def\ifunc@lled#1{\expandafter\ifx\csname @qnnum#1\endcsname\relax}
\let\@qnend=\end\gdef\end{\if@qnfile
\immediate\write16{Equation numbers written on []\jobname.EQN.}\fi\@qnend}
\newcount\r@fcount \r@fcount=0
\newcount\r@fcurr
\immediate\newwrite\reffile
\newif\ifr@ffile\r@ffilefalse
\def\w@rnwrite#1{\ifr@ffile\immediate\write\reffile{#1}\fi\message{#1}}
\def\writer@f#1>>{}
\def\referencefile{
  \r@ffiletrue\immediate\openout\reffile=\jobname.ref%
  \def\writer@f##1>>{\ifr@ffile\immediate\write\reffile%
    {\noexpand\refis{##1} = \csname r@fnum##1\endcsname = %
     \expandafter\expandafter\expandafter\strip@t\expandafter%
     \meaning\csname r@ftext\csname r@fnum##1\endcsname\endcsname}\fi}%
  \def\strip@t##1>>{}}

\def\citeall#1{\xdef#1##1{#1{\noexpand\cite{##1}}}}
\def\cite#1{\each@rg\citer@nge{#1}}	
\def\each@rg#1#2{{\let\thecsname=#1\expandafter\first@rg#2,\end,}}
\def\first@rg#1,{\thecsname{#1}\apply@rg}	
\def\apply@rg#1,{\ifx\end#1\let\next=\relax
\else,\thecsname{#1}\let\next=\apply@rg\fi\next}
\def\citer@nge#1{\citedor@nge#1-\end-}	
\def\citer@ngeat#1\end-{#1}
\def\citedor@nge#1-#2-{\ifx\end#2\r@featspace#1 
  \else\citel@@p{#1}{#2}\citer@ngeat\fi}	
\def\citel@@p#1#2{\ifnum#1>#2{\errmessage{Reference range #1-#2\space is bad.}%
    \errhelp{If you cite a series of references by the notation M-N, then M and
    N must be integers, and N must be greater than or equal to M.}}\else%
 {\count0=#1\count1=#2\advance\count1
by1\relax\expandafter\r@fcite\the\count0,%
  \loop\advance\count0 by1\relax
    \ifnum\count0<\count1,\expandafter\r@fcite\the\count0,%
  \repeat}\fi}
\def\r@featspace#1#2 {\r@fcite#1#2,}	
\def\r@fcite#1,{\ifuncit@d{#1}
    \newr@f{#1}%
    \expandafter\gdef\csname r@ftext\number\r@fcount\endcsname%
                     {\message{Reference #1 to be supplied.}%
                      \writer@f#1>>#1 to be supplied.\par}%
 \fi%
 \csname r@fnum#1\endcsname}
\def\ifuncit@d#1{\expandafter\ifx\csname r@fnum#1\endcsname\relax}%
\def\newr@f#1{\global\advance\r@fcount by1%
    \expandafter\xdef\csname r@fnum#1\endcsname{\number\r@fcount}}
\let\r@fis=\refis			
\def\refis#1#2#3\par{\ifuncit@d{#1}
   \newr@f{#1}%
   \w@rnwrite{Reference #1=\number\r@fcount\space is not cited up to now.}\fi%
  \expandafter\gdef\csname r@ftext\csname r@fnum#1\endcsname\endcsname%
  {\writer@f#1>>#2#3\par}}
\def\ignoreuncited{
   \def\refis##1##2##3\par{\ifuncit@d{##1}%
     \else\expandafter\gdef\csname r@ftext\csname
r@fnum##1\endcsname\endcsname%
     {\writer@f##1>>##2##3\par}\fi}}
\def\r@ferr{\endreferences\errmessage{I was expecting to see
\noexpand\endreferences before now;  I have inserted it here.}}
\let\r@ferences=\references
\def\references{\r@ferences\def\endmode{\r@ferr\par\endgroup}}
\let\endr@ferences=\endreferences
\def\endreferences{\r@fcurr=0
  {\loop\ifnum\r@fcurr<\r@fcount
    \advance\r@fcurr by 1\relax\expandafter\r@fis\expandafter{\number\r@fcurr}%
    \csname r@ftext\number\r@fcurr\endcsname%
  \repeat}\gdef\r@ferr{}\endr@ferences}
\let\r@fend=\endpaper\gdef\endpaper{\ifr@ffile
\immediate\write16{Cross References written on []\jobname.REF.}\fi\r@fend}
\catcode`@=12
\citeall\refto		
\citeall\ref		%
\citeall\Ref		%
\citeall\r		%
\ignoreuncited
\def\frac#1,#2{{#1\over #2}}
\def\fr#1,#2{{#1\over #2}}

\singlespace
\preprintno{SMUHEP/93--19}
\doublespace

\title Schwinger Model in the Light-Cone Representation
\author Gary McCartor
\affil\smu

\abstract\
I present a solution to the Schwinger model in the light-cone
representation which corrects an error in a previous work.  I
emphasize the details of the mechanism by which the physical vacuum is
different than the perturbative vacuum.  I suggest that the method of
analyzing vacuum structure presented here may be of use in more
complicated theories such as QCD.

\endtitlepage
\oneandahalfspace

\head{1. Introduction}
\taghead{1.}

In a previous paper\r{mccartor91} I presented several solutions to the
Schwinger model in light-cone gauge, some quantized at equal time,
some quantized on the characteristic.  The starting point in all cases
was the Coulomb gauge solution of Nakawaki\r{nakawaki83}.  One of the
solutions in reference 1 has an inconsistency.  In the present paper I
shall point out the inconsistency and supply what I believe is a
correct construction.

In presenting the solution we shall pay particular attention to the
mechanism by which degenerate vacua are possible in the light-cone
representation ( in spite of formal arguments which suggest that only
the perturbative vacuum can be a physical vacuum in that
representation ).  The effect is precisely like an anomaly: one is
faced with an ill defined operator product; in giving a precise
definition to that product one cannot maintain all properties of the
classical product; in this case one must either give up gauge
invariance or the kinematical nature of the operator $P^+$ which forms
the basis of the argument that degenerate vacua are not possible in
the light-cone representation.  Even though other states than the
perturbative vacuum do become degenerate with it under the
interaction, and even though the operator $P^+$ does change ( slightly
) due to the interaction, the degenerate physical vacua of the model
are much simpler in the light-cone representation than in the
equal-time representation and the operator $P^+$ is much simpler than
the manifestly dynamical operator $P^-$.  The corrections to $P^+$ are
independent of the coupling constant so the effect is nonperturbative.
If one wishes to find the states degenerate with the vacuum it is only
necessary to study $P^+$ ( a full description of the dynamics
certainly requires the study of $P^-$ ).  Such an effect may be
present in more complicated theories and the study of a fully gauge
invariant and renormalized $P^+$ may provide a way to study vacuum
structure without involving the full dynamics.

In what follows we shall attempt to adhere to the following notation
which is consistent with references 1 and 2:

\noindent\underbar{Coordinates:}
$$
x^0 = t;\qquad x^1 = x;\qquad g^{00} = -g^{11} = 1;\qquad g^{10} = g^{01}
= 0
$$
$$
\gamma^0 = \left(\matrix{0 & 1\cr 1 & 0}\right);\qquad
\gamma^1 = \left(\matrix{0 & 1\cr -1 & 0}\right) ;\qquad
\gamma^5 = \left(\matrix{-1 & 0 \cr 0 & 1}\right)
$$
$$
x^+ = x^0 + x^1 ; \qquad x^- = x^0 - x^1 ;
\qquad g^{+-} = g^{-+} = 2 ; \qquad
g^{++} = g^{--} = 0
$$
$$
g_{+-} = g_{-+} = \ha ; \qquad \gamma^+ = \gamma^0 + \gamma^1 ; \qquad
\gamma^- = \gamma^0 - \gamma^1
$$

\noindent\underbar{Particulars:}
$$
m = {e\over \sqrt{\pi}} \qquad
(e{\rm \ is\ the\ electromagnetic\ coupling\ constant})
$$
$$
k_-(n) = \fr{(n-\ha)\pi},{L} ; \qquad k_+(n) = \fr{(n-\ha)\pi},{L}
$$
$$
p_-(n)= \fr{n\pi},{L} ;\qquad p_+(n)=\fr{m^2 L},{4n\pi}
$$
$$
p_+(n) = \ha \left(p_0(n) + p_1(n)\right) ; \qquad p_-(n) = \ha
\left(p_0(n) - p_1(n)\right)
$$
$$
p^-(n) = 2p_+(n) ; \qquad p^+(n) = 2 p_-(n)
$$
\noindent \underbar{$\psi$-field:}

$\psi_-$ -- first component of $\psi$

$\psi_+$ -- second component of $\psi$

\taghead{2.}
\head{2. Solution}

The solution in reference 1 which has the inconsistency is that given
by equations (3.89) to (3.96) of that paper.  The solution is
canonical on the initial value surface and satisfies the equations of
motion.  Furthermore almost all the operators satisfy the Heisenberg
equations; but there are two operators which do not.  These are the
spurions, $\sigma_+$ and $\sigma_-$ ( these operators are defined in
reference 1 and also below ).  The space-time dependence for the
spurions as given in reference 1 was:
$$
\sigma_+(x) = e^{-i{\sqrt{\pi}\over 4Lm}\left((Q-Q_5)x^-+Qx^+\right)}
\sigma_+(0)e^{-i{\sqrt{\pi}\over 4Lm}\left((Q-Q_5)x^-+Qx^+\right)}
      \eqno(sgpxo)
$$
$$
 \sigma_-(x) = e^{-i{\sqrt\pi\over 4Lm}\left((Q+Q_5)x^+-Q_5x^-\right)}
 \sigma_-(0)e^{-i{\sqrt\pi\over 4Lm}\left((Q+Q_5)x^+-Q_5x^-\right)}
            \eqno(sgmxo)
$$
The terms in the dynamical operators of that solution which do not
commute with the spurions, which we shall call $P_0^+$ and $P_0^-$,
were given as:
$$
   P_0^+ =  {1\over 4Lm^2}(Q^2-2QQ_5)   \eqno(pp0)
$$
$$
   P_0^- = {1\over 4Lm^2}(Q^2+2QQ_5)          \eqno(pm0)
$$
The relevant commutator algebra is:
$$
  [Q,\sigma_+(0)] = -[Q_5,\sigma_+(0)] = -\sigma_+(0)  \eqno(cqps)
$$
$$
  [Q,\sigma_-(0)] = [Q_5,\sigma_-(0)] = -\sigma_-(0)  \eqno(cqms)
$$
It is trivial to use these relations to check that the spurions do not
satisfy the Heisenberg equations.  It might be thought that the
obvious way to proceed would be to modify the dynamical operators to
properly translate the spurions and thus produce a canonical system
satisfying the equations of electrodynamics.  Unfortunately that does
not work for it seems that there are no operators which one can choose
for $P^+$ and $P^-$ that will work.  That such a situation is possible
may be of interest in itself but I will not further discuss the point
here.  Rather we shall now give what we believe to be a correct
construction.

The classical Lagrangian density is:
$$
 {\cal L} = \ha \left(i \bar{\psi} \gamma^{\mu}
\partial_{\mu} \psi - i \partial_{\mu} \bar{\psi} \gamma^{\mu} \psi \right)
- \fr{1},{4} F^{\mu \nu} F_{\mu \nu}
-  A^{\mu} {J^\prime}_{\mu}  \eqno(lag)
$$
The need to use $J^\prime$ is discussed in reference 1; the relation
of $J^\prime$ to $J$ depends on boundary conditions.  For the boundary
conditions we will use below ( periodic observables along the initial
characteristics ) it is:
$$
J^{+\prime} =J^+ - \ha J^+(0) - \ha J^-(0) \eqno(jpp)
$$
$$
J^{-\prime} =J^- - \ha J^+(0) - \ha J^-(0) \eqno(jmp)
$$
Where
$$
             J^\mu = \normord{\bar{\psi} \gamma^{\mu} \psi } \eqno(jmu)
$$
and $J^+(0)$ and $J^-(0)$ are the zero modes in $J^+$ and $J^-$.  We
initialize $\psi_+$ on $x^+=0$ with antiperiodic boundary conditions:
$$
    \psi_+(0,x^-) = {1\over\sqrt{2L}}\sum_{n=1}^\infty
   b(n) e^{-ik_-(n)x^-} +
   d^*(n) e^{ik_-(n)x^-}  \eqno(psip)
$$
and $\psi_-$ on $x^-=0$:
$$
    \psi_-(x^+,0) = {1\over\sqrt{2L}}\sum_{n=1}^\infty
   \beta (n) e^{-ik_+(n)x^+} +    \delta ^*(n) e^{ik_+(n)x^+}  \eqno(psim)
$$
We work in the gauge where $A^+$ is independent of $x^-$; in which
case it is independent of space-time\r{mccartor91}.  The equations of
motion are:
$$
\fr{\partial \psi_+},{\partial x^+} +
i{1\over4}e(A^-\psi_++\psi_+A^-) = 0 \eqno(eom1)
$$
$$
{\del\psi_-\over\del x^-}+i{1\over4}e(A^+\psi_-+\psi_-A^+)=0 \eqno(eom2)
$$
$$
\fr{\partial^2 A^-},{\partial x^{-2}} = -\ha {J^\prime}^+ \eqno(eom3)
$$
$$
\fr{\partial^2 A^-},{\partial x^+ \partial x^-}
= \ha {J^\prime}^-  \eqno(eom4)
$$
The operator solution is most easily written in terms of the fusion
operators which we take to be:
$$
i\sqrt{n} C(n)  = \sum^{n}_{\ell = 1} d\left(\ell\right) b\left(n
- \ell + 1\right) + \sum^{\infty}_{\ell = 1} b^* \left(\ell\right) b
\left(\ell + n \right)
- d^* \left(\ell\right) d
\left(\ell + n \right)  \eqno(fop)
$$
$$
\sqrt{n} D(n)  = \sum^{n}_{\ell = 1} \delta \left(\ell\right)
\beta \left(n
- \ell + 1\right) + \sum^{\infty}_{\ell = 1} \beta^*
\left(\ell\right) \beta
\left(\ell + n \right) - \delta^* \left(\ell\right) \delta
\left(\ell + n \right) \eqno(fom)
$$
The reason for the change in phase between the fusion operators
associated with the $\psi_+$ and $\psi_-$ fields is to produce
agreement with the notation of references 1 and 2.  To define the
spurion operators we first define the set of states $\ket{M,N}$:
$$
\eqalignno{
\ket{M,N}&=\delta^*\Bigl(M\Bigr)\dots \delta^*\Bigl(1\Bigr)
	d^*\Bigl(N\Bigr)\dots d^*\Bigl(1\Bigr)\ket0
\qquad(M>0,N>0)\cr
\ket{M,N}&=\beta^*\Bigl(M\Bigr)\dots \beta^*\Bigl(1\Bigr)
	d^*\Bigl(N\Bigr)\dots d^*\Bigl(1\Bigr)\ket0
\qquad(M<0,N>0)\cr
\ket{M,N}&=\delta^*\Bigl(M\Bigr)\dots \delta^*\Bigl(1\Bigr)
	b^*\Bigl(N\Bigr)\dots b^*\Bigl(1\Bigr)\ket0
\qquad(M>0,N<0)\cr
\ket{M,N}&=\beta^*\Bigl(M\Bigr)\dots \beta^*\Bigl(1\Bigr)
	b^*\Bigl(N\Bigr)\dots b^*\Bigl(1\Bigr)\ket0
\qquad(M<0,N<0) &(vnm)
}
$$
The spurions can then be defined as:
$$
[\sigma_+,D(n)]= [\sigma_+,D^*(n)]=
[\sigma_+,C(n)]= [\sigma_+,C^*(n)]= 0 \eqno(csgd)
$$
$$
[\sigma_-,D(n)]= [\sigma_-,D^*(n)]=
[\sigma_-,C(n)]= [\sigma_-,C^*(n)]= 0 \eqno(csgc)
$$
$$
             (\sigma_+)^I \ket{M,N} = \ket{M,N+I}  \eqno(spnm)
$$
$$
             (\sigma_-)^I \ket{M,N} = \ket{M+I,N}  \eqno(smnm)
$$
We need the charges $Q_+$ and $Q_-$:
$$
Q_+ = 2 e \sum^{\infty}_{n=1} b^* \left(n\right) b
\left(n\right) - d^* \left(n\right) d \left(n\right)  \eqno(qp)
$$
$$
Q_- = 2 e \sum^{\infty}_{n=1} \beta^* \left(n\right) \beta
\left(n\right) - \delta^*
\left(n\right) \delta \left(n\right) \eqno(qm)
$$
These are related to the charge and pseudocharge by:
$$
\eqalignno{
   Q&={1\over2}(Q_++Q_-)\cr
   Q_5&={1\over2}(Q_--Q_+)  &(qq5)
}
$$
Central to the issue we most want to discuss is the definition of
Fermi products.  We take:
$$
\eqalignno{
&\normord{\psi_+^*(x) \psi_+(x)} \equiv \cr
&\quad \lim_{\epsilon^-\rightarrow0} \left(
e^{-ie\int_x^{x+\epsilon^-} A_-^{(-)} dx^-}
\psi_+^*(x+\epsilon^-) \psi_+(x)
e^{-ie\int_x^{x+\epsilon^-} A_-^{(+)}dx^-}
-{\rm V.E.V.}\right) &(psipp)
}
$$
$$
\eqalignno{
&\normord{ \psi_-^*(x) \psi_-(x)} \equiv \cr
&\quad \lim_{\epsilon^+\rightarrow0}
\left( e^{-ie\int_x^{x+\epsilon^+} A_+^{(-)} dx^+}
  \psi_-^*(x+\epsilon^+) \psi_-(x)
e^{-ie\int_x^{x+\epsilon^+} A_+^{(+)} dx^+} -
{\rm V.E.V.}\right) &(psimp)
}
$$
We shall understand the product defining the coupling term in the
Lagrangian as:
$$
\lim_{\epsilon \rightarrow 0\atop \e^2<0}
\ha [A_{\mu} (x + \epsilon) {J^\prime}^{\mu}
(x) + {J^\prime}^{\mu} (x) A_{\mu} (x - \epsilon) ] \;\;\; .
\eqno(e51)
$$

With these definitions we can now give the operator solution as:
$$
\psi_+ = {1\over \sqrt{2L}}
e^{\lambda_+^{(-)}(x)}
\sigma_+(x)
e^{\lambda_+^{(+)}(x)}
					\eqno(real139)
$$
where
$$
\lambda_+(x) = -i\sqrt{{\pi\over L}}\sum_{n=1}^\infty
{1\over \sqrt{p_-(n)}}\left( C(n)e^{-ip(n)x}
+ C^*(n)e^{ip(n)x}\right)      \eqno(real141)
$$
and
$$
\sigma_+(x) = e^{-i{\sqrt{\pi}\over 4Lm}\left(Q_+(x^--x^+)
\right)}
\sigma_+(0)e^{-i{\sqrt{\pi}\over 4Lm}\left(Q_+(x^--x^+)\right)}
      \eqno(real142)
$$

$$
\psi_- = {1\over\sqrt{2L}} e^{-\lambda_D^*(x^+)}
\sigma_-(x)e^{\lambda_D(x^+)}        \eqno(real143)
$$
where
$$
\lambda_D(x^+) = \sum_{n=1}^\infty{1\over\sqrt{n}}
D(n)e^{-ik_+(n)x^+}
					\eqno(real140)
$$
and
$$
\sigma_-(x) = e^{-i{\sqrt\pi\over 4Lm}\left(Q_-(x^+-x^-)\right)}
\sigma_-(0)e^{-i{\sqrt\pi\over 4Lm}\left(Q_-(x^+-x^-)\right)}
            \eqno(real144)
$$

$$
\eqalignno{
A^-&=-{i\over \sqrt{L}m}\sum_{n=1}^\infty{p^-(n)
\over \sqrt{p_-(n)}}\left( C(n)e^{-ip(n)x}-
 C^*(n)e^{ip(n)x}\right) \cr
&\qquad-{1\over Lm^2}Q_+
&(real145)
}
$$
$$
   A^+=-{1\over Lm^2}Q_-        . \eqno(real146)
$$

The dynamical operators are:
$$
P^- = {1\over 4Lm^2}(Q_-^2-Q_+^2)+\sum_{n=1}^\infty
  p^-(n) C^*(n) C(n) +
\sum_{n=1}^\infty 2k_+(n) D^*(n)D(n) .\eqno(real155)
$$
$$
P^+ = {1\over 4Lm^2}(Q_+^2-Q_-^2) + 2
\sum_{n=1}^\infty p_-(n) C^*(n) C(n) .
					\eqno(real152)
$$
These relations can be used to calculate the currents which are:
$$
      J^+ = {im\over \sqrt{L}}\sum_{n=1}^\infty
  \sqrt{p_-(n)} \Biggl(C^*(n)
      e^{ip(n)x}- C(n)e^{-ip(n)x}\Biggr) +
   {1\over L}Q            \eqno(real150)
$$
$$
\eqalignno{
      J^- = {im\over \sqrt{L}}&\sum_{n=1}^\infty
 {p_+(n)\over \sqrt{p_-(n)}} \Biggl(C(n)
      e^{-ip(n)x}- C^*(n)e^{ip(n)x}\Biggr) \cr
 + \fr{e},{L} &\sum_{n=1}^\infty \sqrt{n}
\left(D^*(n)e^{ik_+(n)x^+}+D(n)e^{-ik_+(n)x^+}\right) +
   {1\over L}Q . &(real151)
}
$$
To complete the specification of the solution we must also define a
physical subspace.  It is usual to define the physical subspace for
the Schwinger model to be the charge zero sector and we must do that
here; if $ \ket{p}$ is in the physical subspace then:
$$
                       Q \ket{p} = 0   \eqno(pss1)
$$
Additionally, we must impose:
$$
                       D(n) \ket{p} = 0   \eqno(pss2)
$$
States in the set \(vnm) which have the form $\ket{M,-M}$ are in the
physical subspace and we can choose any linear combination of these
states for the vacuum then generate the entire representation space by
applying polynomials in the $C^*$ and $D^*$ to that vacuum; if we wish
to impose cluster decomposition we must choose a $\theta$-state for
the vacuum \r{ks75}.  It is easy to check that the Heisenberg
equations are all satisfied and that equations \(eom1) - \(eom3) are
satisfied.  Equation \(eom4) is not satisfied.  To examine that fact
in more detail we write out the left hand side and right hand side of
equation \(eom4) explicitly:
$$
\eqalignno{
 &{ie\over 2\sqrt{\pi L}} \sum_{n=1}^\infty
 {p_+(n)\over \sqrt{p_-(n)}} \Biggl(C(n)
      e^{-ip(n)x}- C^*(n)e^{ip(n)x}\Biggr) \cr
&\ne \cr
&{ie\over 2\sqrt{\pi L}} \sum_{n=1}^\infty
 {p_+(n)\over \sqrt{p_-(n)}} \Biggl(C(n)
      e^{-ip(n)x}- C^*(n)e^{ip(n)x}\Biggr) \cr
 +  &\sum_{n=1}^\infty
\left(D^*(n)e^{ik_-(n)x^+}+D(n)e^{-ik_-(n)x^+}\right) &(exeom4)
}
$$
The difficulty is seen to be the last sum on the right hand side,
which involves the $D$'s and $D^*$'s.  That inequality is the reason
we must make the specification of equation \(pss2).  With that
specification of the physical subspace equation \(eom4) is satisfied
in matrix elements between physical states.

The only two complicated issues associated with finding the solution
\(real139) - \(real152) are finding the zero modes in the A-fields and
calculating the dynamical operators; we shall now discuss these two
points.  The boundary conditions we choose require that $A^-$ be
periodic along $x^-$.  That is only possible if equations \(eom3) and
\(eom4) have no zero mode.  Those are the conditions which determine
the zero modes in $A^+$ and $A^-$. Using \(jmu), \(jpp), \(jmp),
\(psipp) and \(psimp) we find that the requirement that \(eom3) and
\(eom4) have no zero modes gives the relations:
$$
    A^-_0 + {\pi\over Le^2}Q_+ = 0  \eqno(amz)
$$
$$
    A^+ + {\pi\over Le^2}Q_- = 0  \eqno(apz)
$$
where $A^-_0$ is the zero mode of $A^-$.

We now turn to the problem of calculating the dynamical operators
$P^-$ and $P^+$.  That these are the correct operators is shown by the
consistency of the solution.  The problem with calculating them occurs
any time different surfaces are used to initialize different fields.
Since the initialization was done on both $x^+ = 0$ and $x^- = 0$ we
expect to have to integrate some density over each surface to
calculate the dynamical operators.  If we work out the formal
densities we find ( see below ) that we need to integrate functionals
of the fields over regions which are not the initial value surfaces
for those fields.  If that were true we would be in the position of
having to solve the problem before we could formulate it.  Generally
one would not expect that situation to arise: if we have properly
chosen a set of degrees of freedom we should be able to calculate the
momenta and energy from the given information.  In all cases where the
answer is known \r{mccartor88} \r{mccartor89} \r{mccartor91} \r{mr94}
the following rule applies: work out the densities as usual but
integrate over a given initial value surface only those parts of these
densities which involve fields initialized on the integration surface;
for gauge theories there is a further consideration which we shall
come to below.  We shall now illustrate this rule and show that it
works in the present case.  Using:
$$
T^{\mu \nu} = \sum_{\phi} \fr{\partial \phi},{\partial x_{\mu}}
\fr{\partial {\cal L}},{\partial (\partial_{\nu} \phi)}
- g^{\mu \nu} {\cal L} \eqno(tmn)
$$
we calculate:
$$
T^{++} = \normord{2i \Bigl(\psi^*_+ \partial_- \psi_+
- \partial_- \psi^*_+
\psi_+\Bigr)}
\eqno(tpp)
$$
$$
T^{+-} = \normord{2i \Bigl(\psi^*_+ \partial_+
\psi_+ - \partial_+ \psi^*_+
\psi_+\Bigr)}
+ \normord{(\partial_- A^-)^2}+2{J^\prime}^\mu A_\mu
\eqno(tpm)
$$
$$
T^{-+} = \normord{2i \Bigl(\psi^*_- \partial_-
\psi_- - \partial_- \psi^*_-
\psi_-\Bigr)} - \normord{(\partial_- A^-)^2}
  +2{J^\prime}^\mu A_\mu
\eqno(tmp)
$$
$$
T^{--}=\normord{2i(\psi^*_-\del_+\psi_--
\del_+\psi^*_-\psi_-)+
	2\del_-A^-\del_+A^-}
					\eqno(tmm)
$$
Using the rule stated above we calculate the dynamical operators as:
$$
     P^- = \ha \int^L_{-L}  - \normord{(\partial_- A^-)^2}
  +2{J^\prime}^\mu A_\mu dx^- +
\ha \int^L_{-L} \normord{2i(\psi^*_-\del_+\psi_--
\del_+\psi^*_-\psi_-)} dx^+
$$
$$
     P^+ =  \ha \int^L_{-L} \normord{2i \Bigl(\psi^*_+
\partial_- \psi_+ - \partial_- \psi^*_+
\psi_+\Bigr)} dx^-
$$
There is a further point associated with the fact we are dealing with
a gauge theory: the fermi products in the above expressions need to be
gauge corrected and for the case of $P^-$ that again leads us to the
need for a field, $A^-$ along $x^- = 0$, where we do not know it; we
again use only what we know and gauge correct only with the (
space-time independent ) zero mode of $A^-$.  We thus calculate the
dynamical operators as given by \(real155), \(real152).

\taghead{3.}
\head{3. Discussion}

Equation \(pss2) removes most of the states associated with the
$\psi_-$ field from the physical subspace.  Indeed the only operators
from that field which remain physical under the interaction are the
spurion, $\sigma_-$, and the charge,$Q_-$.  The $D$-field is an
auxiliary field.  While this may seem strange, the same conclusion can
be reached by gauge transforming the Coulomb gauge solution
\r{nakawaki83} to light-cone gauge \r{mccartor91}.  In the Coulomb
gauge solution only physical degrees of freedom are present; under the
gauge transformation only $\sigma_-$ and $Q_-$ survive from $\psi_-$.
The $D$-field must be added to represent the operator solution.  The
need for auxiliary fields seems to be ubiquitous in light-cone gauge.
Even free Maxwell theory requires one \r{bns91} \r{mr94}.  The present
situation is somewhat different however.  Usually the additional
fields are ghosts and their unphysical nature is manifest in the
commutation relations; here, the $D$-field---which is a perfectly
physical field in free theory--- satisfies normal commutation
relations and the only ways I know to find its unphysical nature are
to gauge transform the Coulomb gauge solution or examine the equation
of motion \(eom4).

Finally, we review the way in which states different from the
perturbative vacuum come to mix with it.  The usual argument that that
cannot happen is that the operator,$P^+$ for the interacting theory is
just the $P^+$, $P^+_{FREE}$, for free theory.  Since the physical
vacuum must be an eigenstate of $P^+$ with eigenvalue 0, and since
there is only one such state in free theory, that must be the vacuum
state.  Indeed the density, \(tpp), we used to calculate $P^+$ has the
same form as the free density, but the need to gauge correct the
products, \(psipp), introduces a modification.  For free theory:
$$
     P^+_{FREE} =  {1\over 4Lm^2}Q_+^2 + 2
   \sum_{n=1}^\infty p_-(n) C^*(n) C(n)
$$
For the interacting theory (see \(real152)):
$$
    P^+ = {1\over 4Lm^2}(Q_+^2-Q_-^2) + 2
   \sum_{n=1}^\infty p_-(n) C^*(n) C(n)
$$
The extra term in $P^+$ which allows the degeneracy of the states,
$\ket{M,N}$, with the perturbative vacuum is independent of the
coupling constant.  Note that the additional term in $P^+$ is composed
of operators associated with the field initialized on the surface $x^-
= 0$.  It seems that that must be the case since for all the degrees
of freedom initialized on $x^+ = 0$ translations by $P^+$ move within
the initial value surface and are thus set by the initial conditions.
For the Schwinger model one can find all the states which are
candidates to mix with the vacuum by studying the operator $P^+$,
which, although not exactly the free $P^+$ is much simpler than $P^-$.
To find the vacuum one would have to apply $P^-$ to the candidate
states but that is a much simpler task than finding the eigenvectors
of $P^-$.  The procedure may provide a useful way to study vacuum
structure even for more complicated theories like QCD.  I hope to
report more on that possibility in the future.

\head{ACKNOWLEDGEMENTS}

It is a pleasure to acknowledge many usful discussions with Prof. Yuji
Nakawaki who discovered the error in reference 1. This work was
supported in part by the U.S. Department of Energy under grant no.
DE-FG05-92ER40722.

\vfill\eject
\references

\refis{bns91}A. Bassetto, G. Nardelli, and R. Soldati,
	{\it Yang-Mills Theories in Algebraic Non-Covariant Gauges}
	(World Scientific, 1991)

\refis{ks75}J.B. Kogut and L. Susskind, \pr, D11, 1975, 3594,

\refis{mccartor88}G. McCartor, \zphys, C41, 1988, 271,

\refis{mccartor89}G. McCartor, in {\it Nuclear and Particle
	Physics on the Light Cone,} M.B. Johnson and L.S. Kisslinger,
	eds. (World Scientific, 1988)

\refis{mccartor91}G. McCartor, \zphys, C52, 1991, 611,

\refis{mr94}G. McCartor and D.G. Robertson, {\it Light-Cone
	Quantization of Gauge Fields,} SMU preprint SMUHEP/93--20, {\sl Z.
	Phys.\bf C} in press

\refis{nakawaki83}Y. Nakawaki, \ptp, 70, 1983, 1105,

\endreferences
\endit
\end